\documentclass[preprint,showpacs,preprintnumbers,amsmath,amssymb,aps]{revtex4}

\usepackage{graphicx}
\usepackage{dcolumn}
\usepackage{bm}
\usepackage[dvips]{color}

\def\5-3{Mn$_{5}$Ge$_{3}$}
\def\Ge-Mn{Ge$_{1-x}$Mn$_x$}

\begin{document}

\title{Spin glass-like behavior of Ge:Mn}

\author{C. Jaeger}
  \email{christian.jaeger@wsi.tum.de}
\author{C. Bihler}
\author{T. Vallaitis}
    \altaffiliation[Present address: ]{Universit\"{a}t Karlsruhe, Engesserstr.~5, 76131 Karlsruhe, Germany}
\affiliation{
Walter Schottky Institut, Technische Universit\"{a}t M\"{u}nchen\\
Am Coulombwall 3, 85748 Garching, Germany
}

\author{S. T. B. Goennenwein}
\author{M. Opel}
\author{R. Gross}
 
\affiliation{
Walther-Meissner-Institut, Bayerische Akademie der Wissenschaften\\
Walther-Meissner-Str.~8, 85748 Garching, Germany
}

\author{M. S. Brandt}
\affiliation{
Walter Schottky Institut, Technische Universit\"{a}t M\"{u}nchen\\
Am Coulombwall 3, 85748 Garching, Germany}

\begin{abstract}
We present a detailed study of the magnetic properties of low-temperature molecular beam epitaxy grown Ge:Mn dilute magnetic semiconductor films. We find strong indications for a {frozen state of $\rm Ge_{1-x}\rm Mn_{x}$, with freezing temperatures of $T_f=12$~K and $T_f=15$~K for samples with $x=0.04$ and $x=0.2$, respectively, determined from the difference between field cooled and zero-field cooled magnetization. For $\rm Ge_{0.96}\rm Mn_{0.04}$}, AC susceptibility measurements show a peak around $T_f$, with the peak position $T'_f$ shifting as a function of the driving frequency $f$ by $\Delta T_f'/[T_f'\cdot \Delta$log$f]\approx 0.06${, whereas for sample $\rm Ge_{0.8}\rm Mn_{0.2}$ a more complicated behavior is observed}. Furthermore, {both samples exhibit} relaxation effects of the magnetization after switching the magnitude of the external magnetic field below $T_f$ which are in qualitative agreement with the field and zero-field cooled magnetization measurements. These findings consistently show that Ge:Mn exhibits a {frozen magnetic state} at low temperatures, and that it is not a conventional ferromagnet.

\end{abstract}

\maketitle

\section{\label{sec:introduction}INTRODUCTION}

Diluted magnetic semiconductors (DMS) - obtained by doping semiconductor materials with magnetic impurities - have been investigated extensively due to their potential application in spintronic devices. In particular, the most prominent material system $\text{Ga}_{1-x}\text{Mn}_x\text{As}$ is generally assumed to exhibit a hole-mediated long-range ferromagnetic order with Curie temperatures of up to $T_C = 172~\text{K}$.\cite{Edmonds2004, Nazmul2003, Chiba2003} 
In contrast to conventional ferromagnetism, many DMS systems were found to exhibit spin glass state. As an example, the observation of a spin glass phase in Mn-doped II-VI DMSs like $\text{Zn}_{1-x}\text{Mn}_x\text{Te}$,\cite{McAlister1984} $\text{Cd}_{1-x}\text{Mn}_x\text{Te}$,\cite{Galazka1980, Oseroff1982} and $\text{Cd}_{1-x}\text{Mn}_x\text{Se}$ ($x>0.2$)\cite{Oseroff1982} has been reported more than two decades ago. Recently, spin glass behavior was discovered for the III-V diluted magnetic semiconductors $\text{Ga}_{1-x}\text{Mn}_x\text{N}$ with a spin freezing temperature of $T_f=4.5~\rm K$ ($x\approx0.1$)\cite{Dhar2003} and Te-doped $\text{Ga}_{1-x}\text{Mn}_x\text{As}$ with $T_f=30~\rm K$ ($x=0.085$).\cite{Yuldashev2004}

As the first ferromagnetic group-IV DMS, Park \textit{et al.} reported in 2002 the growth of \Ge-Mn with $T_C$ up to $116~\rm K$ for $x=0.033$.\cite{Park2002} The control of the hole densities in gated Hall bar samples allowed a switching of the ferromagnetic order, which was used as a proof of hole-mediated ferromagnetism in this material. {Transmission electron microscopy (TEM) measurements by Park \textit{et al.} revealed Mn phase separation in small precipitates (2 to 6~nm in diameter) with higher Mn concentration (10\% to 15\%) than the surrounding matrix.\cite{Park2002}} Cho \textit{et al.} reported ferromagnetism with a $T_C=285~\rm K$ in Mn-doped bulk single crystals,\cite{Cho2002} but the magnetic properties of their samples were clearly dominated by the presence of the intermetallic compound $\rm Mn_{11}\rm Ge_8$.\cite{Yamada1986, Kang2005} These reports of high-$T_C$ ferromagnetism in \Ge-Mn are in contrast to recent findings of Li \textit{et al.},\cite{Li2005, Li2005a} who propose to use two ordering temperatures ($T_C^*$ and $T_C$) to describe the magnetic coupling in Ge:Mn. Here, the higher transition temperature $T_C^*$ refers to the onset of local ferromagnetism, whereas only at a much lower transition temperature $T_C$ a percolation transition leading to global ferromagnetism takes place. The values found for $T_C$ for a sample with $5 ~\rm at.\%$ Mn are of the order of $10~\rm K$, which indicates that \Ge-Mn is far away from being a high-$T_C$ DMS. Also, scanning photoelectron microscopy analysis by Kang \textit{et al.} indicates that ferromagnetism in Mn-doped Ge is not of intrinsic nature,\cite{Kang2005} but arises from magnetic properties of Mn-rich clusters in phase-segregated \Ge-Mn. {Similarly, Sugahara \textit{et al.}\cite{Sug05} reported precipitation of amorphous \Ge-Mn clusters as the origin of ferromagnetism in epitaxially grown Mn-doped Ge films.}

In this work, we present a detailed study of the magnetic properties of low-temperature molecular beam epitaxy (LT-MBE) grown Ge:Mn films. {Magnetization measurements for different cooling fields indicate the presence of two different magnetic phases: (1) superparamagnetic \5-3 clusters undergoing a blocking transition around $T_b=210$~K for $\rm Ge_{0.96} \rm Mn_{0.04}$ and $T_b=270$~K for $\rm Ge_{0.8} \rm Mn_{0.2}$, and (2) superparamagnetic Mn-rich nanoclusters performing a blocking transition around $T_f=12$~K for $\rm Ge_{0.96} \rm Mn_{0.04}$ and $T_f=15$~K for $\rm Ge_{0.8} \rm Mn_{0.2}$. To differenciate the different types of clusters, we will consistently call them \5-3 clusters and (Mn-rich) nanoclusters, respectively. In this paper, we concentrate on the transition taking place at $T_f$. For temperatures around $T_f$, we observe a} frequency dependent shift of the AC susceptibility peak, and relaxation effects in time-dependent magnetization measurements{, which} are strong indications for the presence of {inter-nanocluster interactions} at low temperatures. These findings suggest a more complex magnetic behavior of Ge:Mn, {different from} conventional ferromagnetism.

\section{\label{sec:experiment}EXPERIMENTAL DETAILS}

The Ge:Mn samples studied were grown on Ge(100) substrates via LT-MBE. Prior to the deposition process, the substrates were heated to $600^\circ$C for 30 minutes in the MBE system to evaporate the oxide layer. The flux from the Mn effusion cell was calibrated by elastic recoil detection (ERD) analysis and energy dispersive x-ray (EDX) measurements. We investigated growth at different substrate temperatures ($110^\circ$C $\leq T_S\leq 225^\circ$C) and growth rates ($0.1$ \AA/s $\leq R_{\rm Ge} \leq 1$ \AA/s) and discuss the magnetic properties of two samples in detail in this work characteristic for two different ranges of Mn concentration with a composition of $\rm Ge_{0.96} \rm Mn_{0.04}$ ($T_S=150^\circ$C, $R_{\rm Ge}=0.3$ \AA/s) and $\rm Ge_{0.8} \rm Mn_{0.2}$ ($T_S=225^\circ$C, $R_{\rm Ge}=1$ \AA/s).

All magnetization measurements were performed using a superconducting quantum interference device (SQUID) magnetometer. The temperature dependence of magnetization $M(T)$ was measured between $2~\rm K$ and $330~\rm K$ warming-up the sample in a constant external magnetic field $\mu_0H_m=1~\rm mT$ [indicated by solid symbols in the figures below] and $\mu_0H_m=100~\rm mT$ [open symbols]. All fields were applied in the film plane. To investigate the influence of the thermal history of the samples on the measured magnetization curves, we applied different cooling procedures. In the following, cooling the sample in the maximum available magnetic field of $\mu_0H_C=7~\rm T$ is called maximum field cooling (MFC), whereas cooling without any applied magnetic field (nominal $\mu_0H_C=0~\rm T$) is denoted as zero-field cooling (ZFC). For the field cooled (FC) measurement, the measuring field is identical to the field applied during cooling the sample. 

\section{\label{sec:results}RESULTS AND DISCUSSION}

\subsection{Sample $\rm \bf Ge_{0.96} \rm  \bf Mn_{0.04}$}

\begin{figure}[t]
	\centering
		\includegraphics[width=.48\textwidth]{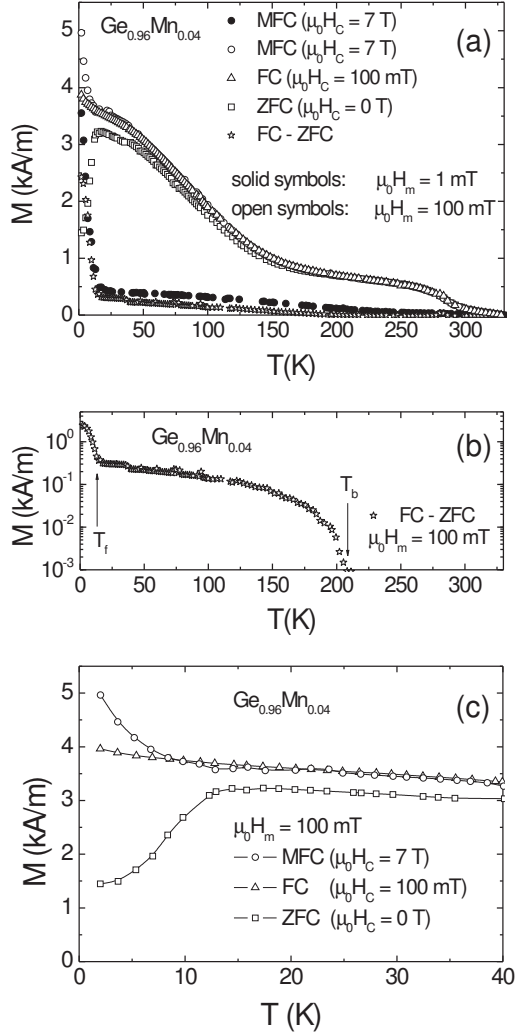}
	\caption{(a) Temperature dependence of the magnetization of the $\rm Ge_{0.96} \rm Mn_{0.04}$ sample. The measurements were performed at $\mu_0H_m=1~\rm mT$ (solid symbols) and $\mu_0H_m=100~\rm mT$ (open symbols). The sample was cooled down in the maximum external magnetic field (MFC, $\mu_0H_C=7~\rm T$), in zero magnetic field (ZFC) and in a field identical to the measuring field (FC). The open stars denote the difference between FC and ZFC magnetization. (b) Logarithmic plot of FC-ZFC difference. (c) Magnification of the low temperature regime of the measurements with $\mu_0H_m=100~\rm mT$ in (a).}
	\label{fig:FIG.1}
\end{figure}

Figure~\ref{fig:FIG.1} shows the temperature dependence of the magnetization of sample $\rm Ge_{0.96} \rm Mn_{0.04}$.  {We performed MFC, FC, and ZFC measurements at $\mu_0H_m=100~\rm mT$ [open symbols], as well as a MFC measurement at $\mu_0H_m=1~\rm mT$ [solid symbols]. Inspecting the difference of the FC and ZFC measurements [open stars, Fig.~\ref{fig:FIG.1}(a),(b)], we observe two temperatures, $T_f=12$~K and $T_b\approx210$~K, below which FC and ZFC magnetizations start to differ.}

{Such a FC-ZFC difference is often interpreted as a fingerprint for spin glass systems.\cite{Dhar2003, Furdyna1988} Another explanation is provided by a blocking transition of superparamagnetic particles. According to the N${\rm \acute{e}}$el-Brown model,\cite{Neel1949, Brown1963} the magnetization aligns along the magnetic easy direction of magnetic anisotropy below the blocking temperature. For a reorientation of magnetization by 180$^\circ$, an energy barrier $E_B$ has to be overcome. Therefore, below its blocking temperature the relaxation of each particle is governed by its $E_B$. The presence of weak to medium interparticle (\textit{e.g.} dipolar) interactions results in a spread of the individual $E_B$s resulting in inhomogeneous freezing.\cite{Dormann1999} For stronger interactions, it is not possible to identify individual $E_B$s anymore, only the average energy of the particle ensemble is relevant and a so-called collective state is present (homogeneous freezing).\cite{Dormann1999} Depending on the interparticle interaction strentgh, it can be difficult to distinguish between an interacting superparamagnet and a real spin glass showing a thermodynamic phase transition.}

\begin{figure}[t]
	\centering
		\includegraphics[width=.5\textwidth]{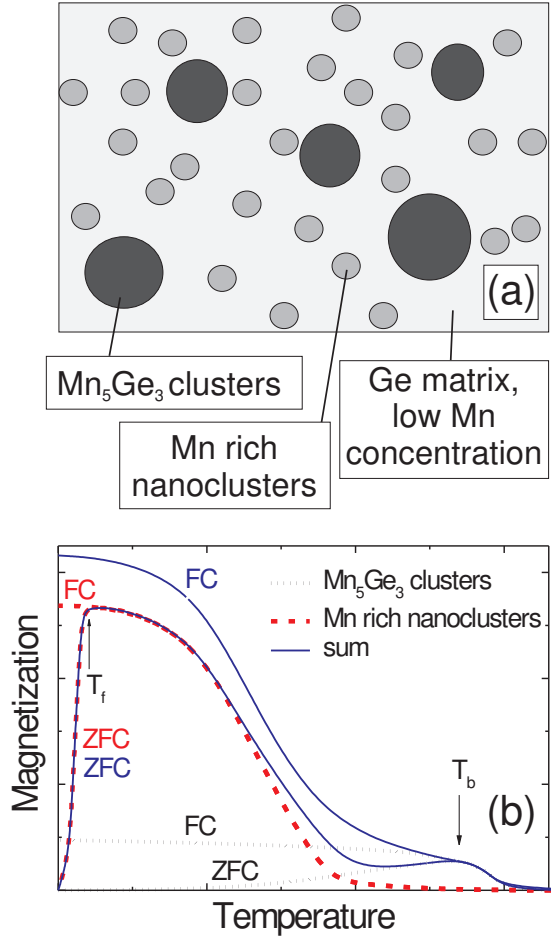}
	\caption{[Color online] (a) Illustration of the two different kinds of clusters present in our samples. (b) Schematic FC and ZFC magnetization curves for both kinds of clusters [black dotted curves for the contribution of the \5-3 clusters, red dashed curves for the Mn-rich nanoparticles] and the sum of both [solid blue curves].}
	\label{fig:Cluster}
\end{figure}

{The fact that we observe two transition temperatures $T_f$ and $T_b$, we attribute to blocking or freezing transitions of two different kinds of superparamagnetic precipitates $-$ Mn-rich nanoclusters and \5-3 clusters, respectively, which are schematically illustrated in Fig.~\ref{fig:Cluster}(a). In Figure~\ref{fig:Cluster}(b) the expected FC and ZFC magnetization curves for the Mn-rich nanoclusters [red dashed curves] and the \5-3 clusters [black dotted curves] are plotted schematically. The sum of both FC and the sum of both ZFC contributions [blue solid curves in Fig.~\ref{fig:Cluster}(b)] qualitatively explain the experimentally observed FC and ZFC measurements. In the following subsections, the transitions taking place at $T_b$ and $T_f$ will be discussed in more detail.}

\subsubsection{Blocking of superparamagnetic \5-3 clusters at $T_b$}

{As already mentioned, the difference between FC and ZFC measurement below $T_b$ we attribute to a blocking of superparamagnetic \5-3 clusters in this temperature range. The incorporation of these \5-3 clusters in the surrounding Ge:Mn matrix in our films has already been discussed elsewhere.\cite{Bihler2006} In the sample with $x\approx0.03$ discussed in detail there, the \5-3 precipitates are preferentially incorporated with their easy magnetic $[1000]$ axis aligned parallel to the [100] growth direction. Due to the lower growth temperature $T_S=150^\circ$C of sample $\rm Ge_{0.96} \rm Mn_{0.04}$ discussed here, compared to the sample $\rm Ge_{0.97} \rm Mn_{0.03}$ reported on in Ref.~\cite{Bihler2006} ($T_S=225^\circ$C), the average \5-3 cluster diameter $\approx15$~nm in Ref.~\cite{Bihler2006} can be regarded as an upper limit for sample $\rm Ge_{0.96} \rm Mn_{0.04}$, which lies well below typical values for the critical diameter below which each cluster is expected to exhibit a single domain ($\approx15$ to 30~nm).}

{Above $T_b$, the \5-3 clusters show a behavior consistent with superparamagnetism with an increase of magnetic response for increasing measurement field ($\mu_0H_m=1~\rm mT$ and $\mu_0H_m=100~\rm mT$). The magnetization of all measurements performed at $\mu_0H_m=100~\rm mT$ strongly decreases around 300~K, which is in agreement with the Curie temperature $T_C=296$~K of bulk \5-3 reported by Yamada.\cite{Yamada1990} We want to state, that we can not exclude the presence of inter-\5-3-cluster interactions, since no AC susceptibility measurements, or measurements of the time dependence of magnetization were performed in this range of temperature. The focus of this manuscript is on the transition taking place at $T_f$, which will be discussed in detail in the following section.}

\subsubsection{Freezing transition at $T_f$}

\textit{Temperature dependence of magnetization}

In the MFC measurement with $\mu_0H_m=1~\rm mT$, we observe a steep decrease of the magnetization for increasing temperatures below $T_f$. For the MFC measurement at higher $\mu_0H_m=100~\rm mT$, {there is only a small decrease of $M(T)$ below $T_f$ followed by} a shoulder below $\approx150$~K in the $M(T)$ diagram. The fact that the shoulder only appears in the measurement with the higher magnetic field indicates superparamagnetic behavior in this temperature range as well. {However, the temperature range in which the shoulder is observed lies well below the blocking temperature of the \5-3 clusters $T_b\approx210$~K. Therefore, the superparamagnetic response below $\approx 150$~K can not be caused by the \5-3 clusters. We rather attribute it to the presence of Mn-rich nanoclusters already introduced above [Fig.~\ref{fig:Cluster}(a)].}

The formation of such regions of locally increased Mn concentration has already been proposed theoretically for ${\rm Ga}_{1-x}$Mn$_x$As by Timm \textit{et al.}\cite{Timm2002} Within these regions, the holes could be regarded as delocalized leading to ferromagnetic coupling at sufficiently low temperatures.\cite{Kaminski2003} Indeed, Park \textit{et al.}\cite{Park2002} report the observation of Mn-rich ($x \approx 0.10-0.15$) nanoclusters via transmission electron microscopy. {Recently, Sugahara \textit{et al.}~\cite{Sug05} reported the observation of amorphous nanoclusters in epitaxially grown Mn-doped Ge films ($x=0.01$ to 0.06) without precipitation of intermetallic compounds such as \5-3 by high-resolution transmission electron microscopy. These authors note that the amorphous clusters are only visible if the thickness of the TEM specimens (with respect to the projection direction) is comparable to the nanocluster diameter $\approx 5$~nm. For thicker samples the TEM image only shows the diamond-type lattice image of the surrounding matrix.\cite{Sug05} From EDX spectroscopy, they determined a Mn concentration in the nanoclusters ranging from 10\% to 20\%, while the Mn concentration of the surrounding matrix was under the detection limit of EDX. Furthermore, Sugahara \textit{et al.}\cite{Sug05} state that their findings are consistent with the experimental results previously reported by Park \textit{et al.}\cite{Park2002} Since EDX measurements on our samples show a significant amount of Mn atoms in the material surrounding the \5-3 clusters ($x=0.02$ for $\rm Ge_{0.97} \rm Mn_{0.03}$ in Ref.~\cite{Bihler2006}), a clustering of the Mn atoms in regions with higher Mn concentration as described is also highly likely in our samples, causing the superparamagnetic response observed in Fig.~\ref{fig:FIG.1}(a) below $\approx 150$~K. Therefore, the sample consists of clusters of the intermetallic phase \5-3 and of clusters of Mn-rich Ge in a crystalline Ge matrix, as illustrated in Fig.~\ref{fig:Cluster}(a). Most recently, the presence of both \5-3 and amorphous Mn-rich nanoclusters has been observed together in one sample by Passacantando \textit{et al.} via TEM.\cite{Passacantando2006}}

Li \textit{et al.}~\cite{Li2005, Li2005a} explained similar magnetization vs. temperature curves {showing no signature of the presence of \5-3 precipitates} via the picture of percolating bound magnetic polarons (BMPs)~\cite{Kaminski2002, Kaminski2003}. From a Curie-Weiss plot of measurements performed at $100~\rm mT$, they deduce a temperature $T_C^*$, at which the BMPs start forming, while they assign $T_C$ to the end of the steep decrease in $M(T)$ at lower temperatures. Applying the same analysis to our measurements, we obtain transition temperatures of $T_C \approx T_f$ and $T_C^*=83~\rm K$ for $\rm Ge_{0.96} \rm Mn_{0.04}$, in good agreement with the values obtained by Li \textit{et al.}~\cite{Li2005, Li2005a}  Li \textit{et al.} conjecture that the physical Mn-rich nanoclusters could be viewed as a generalization of BMPs.\cite{Li2005a}

{The FC-ZFC difference below $T_f$ [Fig.~\ref{fig:FIG.1}] could be explained by a superparamagnetic blocking transition of these Ge:Mn nanoclusters in the same way as the FC-ZFC difference is a signature of the blocking of the \5-3 clusters around $T_b$.} However, measurements of the AC susceptibility and the time dependence of the magnetization discussed below indicate the presence of a frozen, spin glass-like state {$-$ either real spin glass (homogeneous freezing) or interacting superparamagnets (inhomogeneous freezing)~\cite{Dormann1999_2} $-$} at low temperatures. {Independent of the real physical nature of the low temperature state, all measurements performed suggest a more complex magnetic behaviour of Ge:Mn rather than conventional ferromagnetism.}

\textit{AC susceptibility}

\begin{figure}[t]
	\centering
		\includegraphics[width=0.6\textwidth]{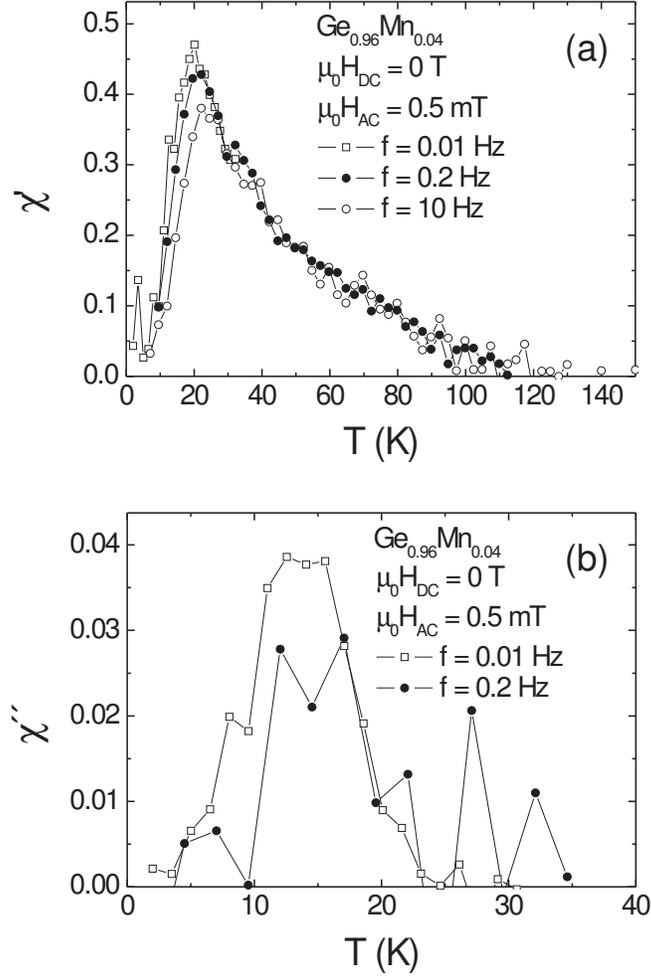}
	\caption{(a) Real part $\chi'(T)$ of the AC susceptibility of the $\rm Ge_{0.96} \rm Mn_{0.04}$ sample. The measurement was performed  with $\mu_0H_{DC}=0~\rm T$ and $\mu_0H_{AC}=0.5~\rm mT$ at different driving frequencies $f=0.01~\rm Hz$ (open squares), $f=0.2~\rm Hz$ (solid circles), and $f=10~\rm Hz$ (open circles). (b) Imaginary part $\chi''(T)$ of the AC susceptibility.}
	\label{fig:FIG.2}
\end{figure}

To learn more about the low-temperature state and its phase transition, frequency-dependent AC susceptibility measurements were performed. The results are shown in Fig.~\ref{fig:FIG.2}, with $\mu_0H_{DC}=0~\rm T$ and $\mu_0H_{AC}=0.5~\rm mT$ at $f=0.01~\rm Hz$ (open squares), $f=0.2~\rm Hz$ (solid circles) and $f=10~\rm Hz$ (open circles). In the real part of the AC susceptibility $\chi'$ (Fig.~\ref{fig:FIG.2}(a)), a pronounced peak is visible for all three frequencies, accompanied by a monotonic decrease of the susceptibility to $\chi'=0$ at $120~\rm K$. This is the temperature range where the shoulder is observed in the magnetization measurements at $\mu_0H_m=100~\rm mT$. {We also come back to this observation in the discussion of sample $\rm Ge_{0.8} \rm Mn_{0.2}$.} The imaginary part of the susceptibility $\chi''(T)$ (Fig.~\ref{fig:FIG.2}(b)) is about a tenth of the real part $\chi'(T)$, which leads to a reduced signal-to-noise ratio due to small sample volume. Nevertheless, a peak of $\chi''(T)$ can be observed for $f=0.01~\rm Hz$ (open squares) and $f=0.2~\rm Hz$ (solid circles).

\begin{figure}[t]
	\centering
		\includegraphics[width=0.6\textwidth]{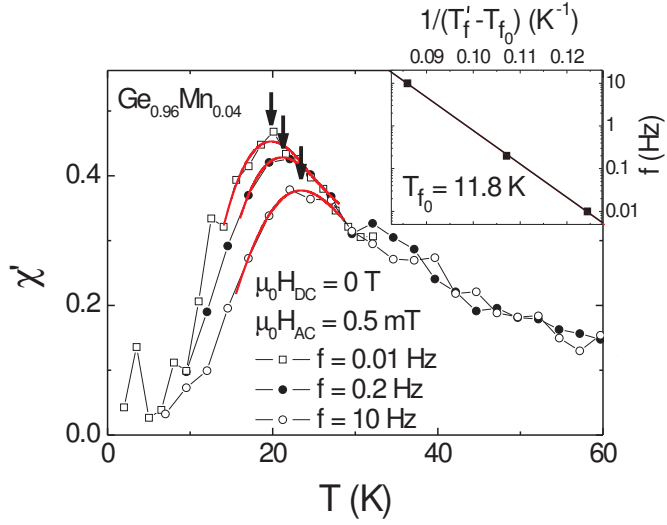}
	\caption{(Color online) Real part $\chi'(T)$ of the AC susceptiblity of the $\rm Ge_{0.96} \rm Mn_{0.04}$ sample in the vicinity of the peak positions. The notation is the same as in Fig.~\ref{fig:FIG.2}. The peak positions $T_f'$ are indicated by arrows, the smooth solid lines are guides to the eye. The inset shows the dependence of the peak position temperatures $T_f'$ (solid squares) on the measuring frequency $f$ fitted with the Vogel-Fulcher law (straight line).}
	\label{fig:FIG.3}
\end{figure}

A careful investigation of $\chi'(T)$ reveals a small shift of the peak position to higher temperatures for higher driving frequencies (Fig.~\ref{fig:FIG.3}). The peak positions are denoted with arrows, the smooth solid lines are guides to the eye. The intensity of the peak increases for lower measuring frequencies. Furthermore, $\chi'(T)$ curves for different measuring frequencies overlap for temperatures higher than the peak temperature. Such a behavior is observed in many spin glass {and disordered magnetic} systems~\cite{Dhar2003, DeTeresa1998, Cardoso2004, Geschwind1988, Dormann1999_2, Dormann1999}. 

A quantitative measure of the frequency shift of the peak position is given by the relative shift of the peak temperature $\Delta T_f'/T_f'$ per decade shift in frequency. For the sample $\rm Ge_{0.96} \rm Mn_{0.04}$ studied here, $C_1=\Delta T_f'/[T_f'\cdot \Delta \text{log} f]\approx 0.06$. {Typical values for} spin glass systems are $C_1=0.02$ for $\text{Cd}_{0.6}\text{Mn}_{0.4}\text{Te}$,\cite{Mauger1988} $C_1=0.05$ for $\text{Eu}_{0.6}\text{Sr}_{0.4}\text{S}$,\cite{Mauger1988} $C_1=0.012$ for $\text{Ga}_{1-x}\text{Mn}_x\text{N}$,\cite{Dhar2003} and $C_1=0.005$ for Cu:Mn.\cite{Mydosh1993}

{For superparamagnetic particles, Dormann \textit{et al.} distinguish three different types of dynamical behavior, depending on the interparticle interaction strength: (1) non-interacting particles for $0.10<C_1<0.13$ (theory), (2) weak interaction regime (inhomogeneous freezing) with $0.03<C_1<0.06$, and (3) medium to strong interaction regime (homogeneous freezing) at $0.005<C_1<0.02$.\cite{Dormann1999} Therefore, the value $C_1=0.06$ found for our sample $\rm Ge_{0.96} \rm Mn_{0.04}$ suggests the presence of at least weak interactions assuming the presence of the superparamagnetic Ge:Mn nanoclusters discussed above.}

Furthermore, the frequency dependence of the peak position of spin glasses {and other disordered magnetic compounds} can be described using the Vogel-Fulcher law~\cite{Tholence1980}
\begin{equation}
f=f_0\cdot \text{exp}\left[-\frac{E_A}{k_B(T_f'-T_{f_0})}\right], 
\end{equation}
with an activation energy $E_A$ and a characteristic frequency $f_0$. Here, $T_{f_0}$ has been interpreted to take into account inter-cluster couplings~\cite{Shtrikman1981}. It can be regarded as the true critical temperature for $f\rightarrow0$, while $T_f'$, being higher than $T_{f_0}$, is the dynamic manifestation of the underlying {freezing} transition~\cite{Mydosh1993}.  
The fit to the three peak positions for sample $\rm Ge_{0.96}\rm Mn_{0.04}$ is shown in the inset of Fig.~\ref{fig:FIG.3} with fit parameters of $E_A/k_B=180~\rm K$, $T_{f_0}=11.8~\rm K$ and $f_0=5\times10^7~\rm Hz$. $T_{f_0}=11.8~\rm K$ is in agreement with the freezing temperature $T_{f}=12~\rm K$ determined above from the difference between the FC and ZFC magnetization measurements. The $E_A/k_B$ values are in the same range as observed for other glassy systems, like $108~\rm K$ for $\text{Fe}_{1/3}\text{TiS}_2$ with $T_{f_0}=48.7~\rm K$~\cite{Koyano1993} and $220~\rm K$ for $\text{Co}_{0.2}\text{Zn}_{0.8}\text{Fe}_{1.6}\text{Ti}_{0.4}\text{O}_4$ with $T_{f_0}=106~\rm K$.\cite{Bhowmik2003} The obtained frequency $f_0$ is of the same order as the observed $f_0\approx10^7~\rm Hz$ for $\text{Co}_{0.2}\text{Zn}_{0.8}\text{Fe}_{1.6}\text{Ti}_{0.4}\text{O}_4$~\cite{Bhowmik2003} and $f_0=2.5\times 10^7~\rm Hz$ for CuMn (with $4.6~\rm  at.\%$ of Mn).\cite{Mydosh1993} 

{A further quantity useful to quantify the frequency dependence of $T_f'$ is $C_2=(T_f'-T_{f_0})/T_f'$. For sample $\rm Ge_{0.96} \rm Mn_{0.04}$, we obtain $C_2=0.4$. For the three different types of dynamical behavior introduced above, Dormann \textit{et al.} distinguish (1) $C_2=1$ for non-interacting particles (theory), (2) $0.3<C_2<0.6$ for the weak interaction regime (inhomogeneous freezing), and (3) $0.07<C_2<0.3$ for the medium to strong interaction regime (homogeneous freezing).\cite{Dormann1999} Therefore, also $C_2=0.4$ of sample $\rm Ge_{0.96} \rm Mn_{0.04}$  suggests the presence of weak interactions between the superparamagnetic nanoclusters, in agreement with the conclusions derived from an analysis of $C_1$.}

\textit{Relaxation effects of magnetization below $T_f$}

\begin{figure}[t]
	\centering
		\includegraphics[width=0.43\textwidth]{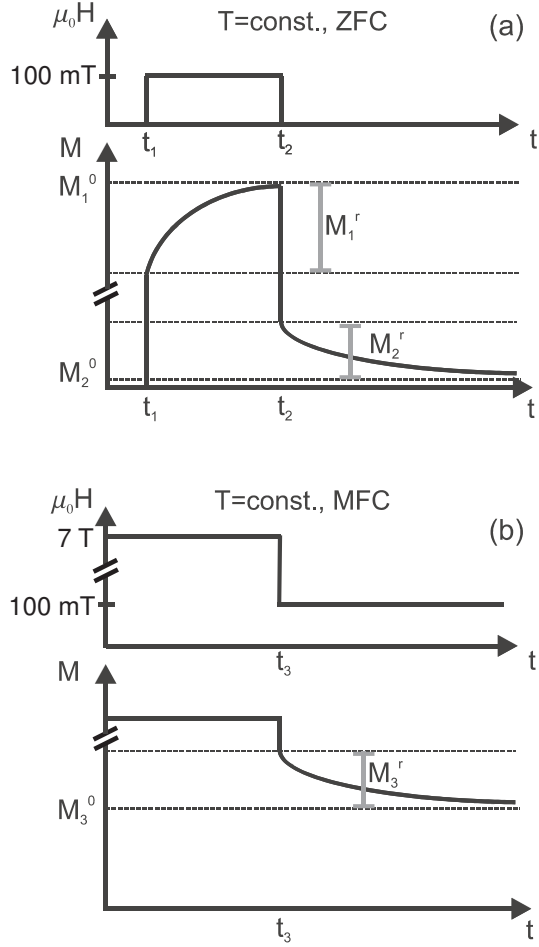}
	\caption{Switching procedures of the external magnetic field (upper panels) for the time-dependent magnetization measurements shown schematically in the lower panels for (a) ZFC and (b) MFC.}
	\label{fig:FIG.4}
\end{figure}

To further elucidate the dynamics of the system, we also performed time-dependent magnetization measurements, using two different measurement procedures displayed schematically in the upper panels of Fig.~\ref{fig:FIG.4}. 

In the first procedure (Fig.~\ref{fig:FIG.4}(a)), the sample was cooled from room temperature to a constant measurement temperature with no external magnetic field applied to the sample (ZFC). After the measurement temperature was stable, the external magnetic field was increased to $\mu_0H=100~\rm mT$ at a time denoted by $t_1$ in Fig.~\ref{fig:FIG.4}(a). After two hours of measurement denoted by $t_2$ in Fig.~\ref{fig:FIG.4}(a), the magnetic field was switched off again. We repeated this procedure for different measurement temperatures of $2~\rm K, 5~\rm K, 10~\rm K, 15~\rm K$, and $40~\rm K$. 

In the second procedure (Fig.~\ref{fig:FIG.4}(b)), the sample was cooled from room temperature to a constant measurement temperature with the maximum field $\mu_0H_C=7~\rm T$ applied to the sample (MFC). Then, we reduced the magnetic field, reaching $100~\rm mT$ at a time $t_3$. This procedure we again performed at different measurement temperatures of $2~\rm K, 5~\rm K, 10~\rm K$, $15~\rm K$, and $40~\rm K$. In the following, the time-dependent magnetization during the time interval $t_1 < t < t_2$ (Fig.~\ref{fig:FIG.4}(a)) will be denoted as $M_1(t)$. Likewise, $M_2(t)$ corresponds to $t>t_2$ (Fig.~\ref{fig:FIG.4}(a)), and $M_3(t)$ describes the results for $t>t_3$ of experiments following the second procedure (Fig.~\ref{fig:FIG.4}(b)).

A schematic illustration of the measured time dependence of the magnetization is shown in the lower panels of Fig.~\ref{fig:FIG.4}. For the ZFC procedure (Fig.~\ref{fig:FIG.4}(a)), the net magnetization in the sample after cooling down is zero. After the magnetic field is switched to $\mu_0H=100~\rm mT$ at $t_1$, the magnetization jumps to a finite value $M_1^0-M_1^r$, followed by an additional slow increase to $M_1^0$. After the magnetic field is switched off again at $t_2$, the magnetization again jumps to a finite value $M_2^0+M_2^r$, followed by a slow decay to $M_2^0$.  
For the MFC procedure (Fig.~\ref{fig:FIG.4}(b)), the sample exhibits a finite magnetization value after cooling down, which is constant in time. After the reduction of the external field to $100~\rm mT$ at $t_3$, the magnetization jumps down to a smaller value $M_3^0+M_3^r$, followed by an additional slow decrease of magnetization with time to $M_3^0$. The jumps most likely correspond to fast relaxation effects, which cannot be resolved due to the finite time (around 100 seconds) required to sweep the external magnetic field.

It is an intrinsic property of glassy systems to react to changes of the magnetic field below its freezing temperature $T_f$ with creeping effects of magnetization.\cite{Rammel1982} This is caused by the fact that the variation of the field creates a nonequilibrium situation. On the other hand, if the field is kept constant (FC) during the cooling below $T_f$, no creeping effects of magnetization are observed. It is important to note that the sketches in Fig.~\ref{fig:FIG.4} are not to scale. The intensities of the creeping effects, which are denoted with $M_1^r$, $M_2^r$ and $M_3^r$ in Fig.~\ref{fig:FIG.4} are generally much smaller than the total magnetization values $M_1^0$, $M_2^0$ and $M_3^0$.  

\begin{figure}[t]
	\centering
		\includegraphics[width=0.35\textwidth]{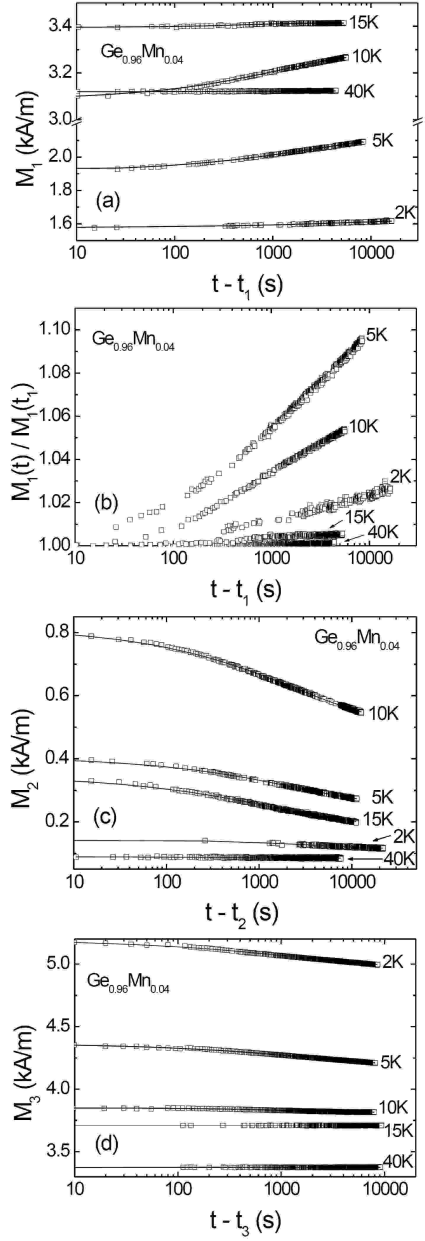}
	\caption{Increase of magnetization $M_1(t)$ (a) and decay of magnetization $M_2(t)$ (c) and $M_3(t)$ (d) measured as a function of time at different constant measurement temperatures. The solid lines are fitted curves. (b) Magnetization $M_1(t)$ normalized to the magnetization $M_1(t_1)$.}
	\label{fig:FIG.5}
\end{figure}

Figure~\ref{fig:FIG.5} (a), (c) and (d) show the magnetization curves $M_1(t)$, $M_2(t)$, and $M_3(t)$ at different temperatures for $\rm Ge_{0.96} \rm Mn_{0.04}$, respectively. In Fig.~\ref{fig:FIG.5}(b), $M_1(t)$ is normalized to the magnetization $M_1(t_1)$ immediately after the field was switched on. The strongest relative increase of magnetization is found for $T=5~\rm K$, whereas for higher and lower temperatures, the relative increase is more moderate.

Many functional forms have been proposed to describe the time dependence of the magnetization in {glassy} systems. Reasonable results are obtained in different systems by fitting with logarithmic~\cite{Dhar2003}, power law~\cite{Kinzel1979}, as well as stretched exponential~\cite{DeFotis1998} time dependencies. The best fit to our data was obtained by using 
\begin{equation}
\label{equ:fit1}
M_1(t)=M_1^0-M_1^r \cdot \text{exp}\left[-\left(\frac{t}{{\tau}}\right)^{1-n}\right],
\end{equation}
and
\begin{equation}
\label{equ:fit2}
M_{2/3}(t)=M_{2/3}^0+M_{2/3}^r \cdot \text{exp}\left[-\left(\frac{t}{\tau}\right)^{1-n}\right], 
\end{equation}
which corresponds to a stretched exponential with an additional constant term $M_i^0$. Such a functional form has been used successfully by different other groups to describe relaxation effects in glassy systems.\cite{Mitchler1993, Maignan1998, Freitas2001} 
Here, the stretched exponential accounts for the glassy contribution to the magnetization, with $\tau$ being a time constant and $n$ affecting the relaxation rate of the glassy component. $M_i^r$ gives the amplitude of the glassy component. The constant term $M_i^0$ is often interpreted as an intrinsic ferromagnetic contribution to the magnetization, which is assumed to be time independent.\cite{Freitas2001} 
The solid lines in Fig.~\ref{fig:FIG.5} are fitted curves using equations (\ref{equ:fit1}) and (\ref{equ:fit2}). The parameter $n$ was found to vary between $0.5$ and $0.6$, except for $M_1(15~{\rm K})$, $M_1(40~{\rm K})$, and $M_3(10~{\rm K})$ which exhibited $n\approx0.2$. Freitas \textit{et al.}\cite{Freitas2001} found values of $0.48<n<0.6$ for the cluster glass material $\rm La_{0.7-x}\rm Y_x \rm Ca_{0.3}\rm MnO_3$. Furthermore, for $M_1$ and $M_2$, $\tau$ decreases in Fig.~\ref{fig:FIG.5} from $\tau \approx 6 \times 10^3$~s at 2~K to $\tau \approx 1 \times 10^3$~s at 40~K. For $M_3$, $\tau$ varies between $1 \times 10^3$~s and $2 \times 10^3$~s.
\begin{figure}[t]
	\centering
		\includegraphics[width=0.6\textwidth]{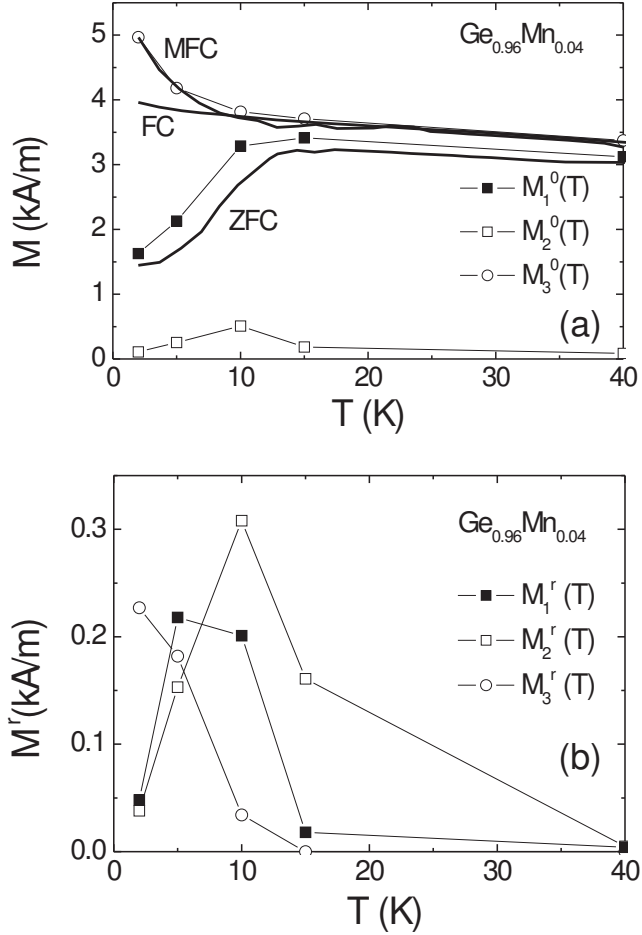}
	\caption{Temperature dependence of fit parameters (a) $M_i^0$  and (b) $M_i^r$ of sample $\rm Ge_{0.96} \rm Mn_{0.04}$. The solid squares display $M_1^0$ and $M_1^r$, the open squares $M_2^0$ and $M_2^r$, and the open circles $M_3^0$ and $M_3^r$. The thick solid lines in (a) are the ZFC, FC and MFC $M(T)$ measurements performed at $\mu_0H_m=100~\rm mT$ (open symbols in Fig.~\ref{fig:FIG.1}).}
	\label{fig:FIG.6}
\end{figure}
Figure~\ref{fig:FIG.6} shows the temperature dependence of the fit parameters $M_0$ and $M_r$. Additionally, the ZFC, FC and MFC $M(T)$ measurements performed at $\mu_0H_m=100~\rm mT$ (open symbols in Fig.~\ref{fig:FIG.1}) are displayed in the graphs as solid lines. 

After switching on the magnetic field at $t_1$, the magnetization jumps to $M^0_1-M^r_1$. Due to the additional relaxation effect $M^r_1$ being much smaller than the time-independent $M^0_1$ (see Fig.~\ref{fig:FIG.6}), the temperature dependence of the magnetization is mainly given by $M^0_1(T)$. $M^0_1(T)$ displayed in Fig.~\ref{fig:FIG.6}(a) by solid squares indeed nicely follows the ZFC magnetization measurement. Therefore, the increase of the ZFC magnetization for increasing temperatures below $T_f$ for the most part has to be an effect of temperature, rather than a relaxation effect in time on the timescale of the $M(T)$ measurement, which takes about 10 minutes from $T=5$~K to 20~K. The same argument is valid for the decrease at low temperatures in the MFC measurement: Since $M^r_3$ is much smaller than the time-independent $M^0_3$ (see Fig.~\ref{fig:FIG.6}), the temperature dependence of the magnetization is mainly given by $M^0_3(T)$, which follows the MFC measurement. Therefore, also the decrease of magnetization in the MFC measurement indeed is a temperature effect.

Furthermore, the temperature dependence of $M_2^0$ shows a maximum around $10~\rm K$ (Fig.~\ref{fig:FIG.6}(a)). This kind of measurement, with a switching of the magnetic field from $\mu_0H=0~\rm T$ to $100~\rm mT$ and back to $0~\rm T$ can be interpreted as half a period of a very slow AC experiment with $f \approx 10^{-4}~\rm Hz$. Therefore, the peak in $M_2^0(T)$ is equivalent to a peak in the AC susceptibility and consequently is correlated to a freezing transition in the sample. The peak temperature $\approx10~\rm K$ nicely coincides with the freezing temperature $T_f=12~\rm K$ deduced from the difference between the FC and ZFC measurements and $T_{f_0}=11.8~\rm K$ obtained from the Vogel-Fulcher analysis.

Further indications for a freezing transition at low temperatures are obtained from the fit parameters $M^r$, which denote the magnitude of the relaxation in the sample. Figure~\ref{fig:FIG.6}(b) shows the different $M^r(T)$, with $M^r_1(T)$ and $M^r_2(T)$ exhibiting a peak around $5~{\rm K}\leq T \leq15~\rm K$. This behavior can be rationalized as follows: {For decreasing temperatures below $T_f$,  $M^r$ decreases as a result of nanocluster freezing, while for increasing temperatures above $T_f$, $M^r$ decreases due to thermal energy exceeding intercluster interactions.}

In summary, also the measurements of the time-dependence of magnetization indicate the presence of a transition to a low temperature frozen state in Ge:Mn, with the transition temperature in complete accordance with the measurements of the temperature-dependence of magnetization and the AC susceptibility presented above.

\subsection{Sample $\rm \bf Ge_{0.8} \rm \bf Mn_{0.2}$}

\begin{figure}[t]
	\centering
		\includegraphics[width=0.35\textwidth]{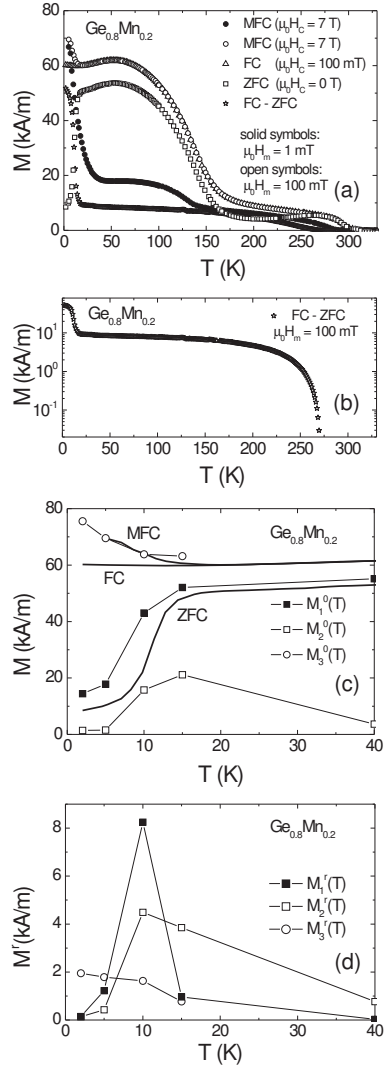}
	\caption{(a) Temperature dependence of magnetization of sample $\rm Ge_{0.8} \rm Mn_{0.2}$. The notation is the same as in Fig.~\ref{fig:FIG.1}. (b) Logarithmic plot of FC-ZFC difference.  (c) and (d) show the temperature dependence of the fit parameters $M^0$ and $M^r$, respectively. The solid squares display $M_1^0$ and $M_1^r$, the open squares $M_2^0$ and $M_2^r$, and the open circles denote $M_3^0$ and $M_3^r$. The thick solid lines in (c) are the ZFC, FC, and MFC $M(T)$ measurements performed at $\mu_0H_m=100~\rm mT$, shown by open symbols in (a).}
	\label{fig:FIG.7}
\end{figure}

{The same experiments and analyses were also performed for sample $\rm Ge_{0.8}\rm Mn_{0.2}$. Figure~\ref{fig:FIG.7}(a) displays the ZFC, FC and MFC curves in $\mu_0H_m=1~\rm mT$ [solid symbols] and $\mu_0H_m=100~\rm mT$ [open symbols]. Again, the difference between the FC and the ZFC measurements [open stars,  Fig.~\ref{fig:FIG.7}(a) and (b)] reveals the presence of two transition temperatures $T_f=15$~K and $T_b=270$~K.}

{Above the blocking temperature of the \5-3 clusters $T_b$, increasing magnetic fields ($\mu_0H_m=1~\rm mT$ and $\mu_0H_m=100~\rm mT$) lead to an increase of magnetization, as observed for superparamagnetic systems. The magnetization strongly decreases above 300~K for all measurements performed at $\mu_0H_m=100~\rm mT$, in agreement with the Curie temperature $T_C=296$~K~\cite{Yamada1990} of bulk \5-3.}
{For sample $\rm Ge_{0.8} \rm Mn_{0.2}$, the superparamagnetic blocking of the \5-3 clusters is also nicely corroborated by the increase of ZFC magnetization for increasing temperature above 200~K, exhibiting a broad maximum around $T_b$ [compare with schematic curves in Fig.~\ref{fig:Cluster}(b)]. For sample $\rm Ge_{0.96} \rm Mn_{0.04}$ this effect is covered by the more intense contribution of the Ge:Mn nanoclusters to the magnetization at these temperatures.}

{In contrast to sample $\rm Ge_{0.96}\rm Mn_{0.04}$, a partial polarization of the Ge:Mn nanoclusters in the sample $\rm Ge_{0.8}\rm Mn_{0.2}$ can already be observed for the MFC measurement in the lower field $\mu_0H_m=1~\rm mT$. For the measurements with $\mu_0H_m=100~\rm mT$, the polarization accordingly increases and the shoulder becomes more pronounced.} 

We want to point out that \textit{in spite of the higher abundance of the \5-3 precipitates evident from the magnetization experiment}, the {freezing behavior at low temperatures} of sample $\rm Ge_{0.8}\rm Mn_{0.2}$ is the same as for the sample $\rm Ge_{0.96}\rm Mn_{0.04}$ with a lower Mn concentration: We again observe a pronounced difference between the FC and ZFC curves with a freezing temperature $T_f=15~\rm K$. Measurements of the time dependence of magnetization of this sample {apart from the amplitudes of the different magnetization components} exhibit the same behavior as already discussed for $\rm Ge_{0.96}\rm Mn_{0.04}$. The fit parameters $M_1^0$ and $M_3^0$ follow the MFC and ZFC curves and a peak in $M_2^0(T)$ is visible around the freezing temperature (see Fig.~\ref{fig:FIG.7}(b)).
Also, the relaxation components $M_1^r(T)$ and $M_3^r(T)$ again show a maximum around $10~\rm K$, which is consistent with a freezing transition in the sample at low temperatures (see Fig.~\ref{fig:FIG.7}(c)).

\begin{figure}[t]
	\centering
		\includegraphics[width=0.6\textwidth]{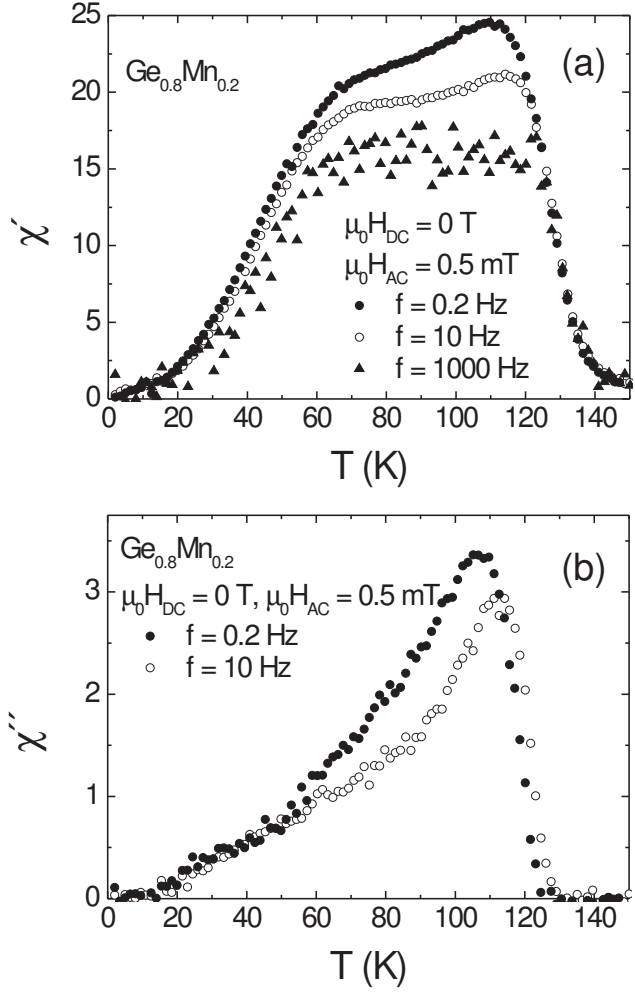}
	\caption{Real part $\chi'(T)$ (a) and imaginary part $\chi''(T)$ (b) of the AC-susceptibility of $\rm Ge_{0.8} \rm Mn_{0.2}$ measured with $\mu_0H_{DC}=0~\rm T$ and $\mu_0H_{AC}=0.5~\rm mT$ at $f=0.2~\rm Hz$ (solid circles), $f=10~\rm Hz$ (open circles) and $f=1000~\rm Hz$ (solid triangles).}
	\label{fig:FIG.8}
\end{figure}

The AC susceptibility measurements performed on this sample (Fig.~\ref{fig:FIG.8}) indicate a more complicated behavior at higher Mn concentration. For all frequencies, $\chi'(T)$ increases strongly between $20~\rm K$ and $60~\rm K$  and subsequently shows a plateau-like behavior up to $115~\rm K$ with slightly increasing $\chi'$ (Fig.~\ref{fig:FIG.8}(a)). Above this peak temperature, the susceptibility decreases rapidly and reaches $\chi'(T)\approx0$ at $140~\rm K$. The imaginary part of the susceptibility $\chi''(T)$ (Fig.~\ref{fig:FIG.8}(b)) increases monotonically at low temperatures and peaks at $T\approx110~\rm K$. For higher temperatures, $\chi''(T)$ decreases strongly and vanishes above 125~K.

{A comparison of the ZFC and $\chi'$ measurements reveals that the plateau-like signal of $\chi'$ spans the same temperature range as the superparamagnetic shoulder in the ZFC curve. The fact, that a field of $\mu_0H_m=1~\rm mT$ in the MFC measurement discussed above is sufficient to achieve a significant polarization of the superparamagnetic nanoclusters easily accounts for the higher signal intensity of the AC susceptibility measurement with the AC field $\mu_0H_{\rm AC}=0.5$~mT in comparison to the behavior observed for the sample $\rm Ge_{0.96}\rm Mn_{0.04}$ with lower Mn concentration. We would expect a Curie temperature for the nanoclusters depending on their respective Mn concentration. From the decrease of $\chi'$ for increasing temperatures above $\approx 115$~K, as well as from magnetization in the MFC, FC, and ZFC measurements, we deduce a characteristic ferromagnetic transition temperature of most of the superparamagnetic nanoclusters in the temperature range 115~K $ \leq T\leq 150$~K. The ferromagnetic phase transition of the nanoclusters might also explain the peak in $\chi''$. A comparison of the shape of the ZFC and $\chi'$ curves of both samples suggests, that for the sample $\rm Ge_{0.96}\rm Mn_{0.04}$ the average Mn concentration of the nanoclusters is shifted to a lower value, accompanied by a decrease of their average Curie temperature.} 

{The onset of the decay of $\chi'$ on the low temperature side of the plateau-like signal we attribute to the freezing temperature $T_f'$ of the AC susceptibility measurement performed with frequency $f$. For decreasing $f$, this decay takes place at decreasing temperatures. Since the temperature dependence of $M^0_2$ (Fig.~\ref{fig:FIG.7}) can also be regarded as a very slow AC experiment ($f\rightarrow0$), the temperature of $\approx 15$~K below which the decay is observed in $M^0_2$, can be assumed to be close to $T_{f_0}$.}

For $\rm Ge_{0.8}\rm Mn_{0.2}$, we observe a good agreement between the $T_f=15~\rm K$ obtained from the FC/ZFC difference and the transition temperature of about $10~\rm K$ from the measurements of the time dependence of magnetization. {The more complicated behavior of AC susceptibility of this sample could be explained by a combination of ferromagnetic transition temperature of the nanoclusters and their frequency dependent freezing.} Consequently, for both samples we observe a strong difference between zero-field cooled and field cooled magnetization below $T_f$, as well as relaxation effects of the magnetization after switching the external magnetic field below $T_f$.

\section{\label{sec:origin}FURTHER POSSIBLE EXPLANATIONS FOR THE SPIN GLASS-LIKE BEHAVIOR BELOW $T_f$} 

The experiments discussed strongly indicate a transition into a frozen state below $T_f$. {The discussion above was carried out considering a superparamagnetic freezing transition of interacting nanoclusers. In this section, we discuss further possible explanations for the observed glassy behavior.}

{Instead of a superparamagnetic freezing transition of interacting nanoclusers, there could also be a spin glass transition of the Ge:Mn matrix at $T_f$.} Assuming the nanoclusters are embedded in this matrix, a freezing of the clusters with random orientation within the matrix below $T_f$ would be expected.\cite{Mydosh1993} {However, the value we determined for $C_1=0.06$ ($x=0.04$), seems to be too high in comparison with real spin glass systems.}

Alternatively, the freezing transition might also occur inside the nanoparticles themselves. For \Ge-Mn, Zhao \textit{et al.}\cite{Zhao2003:01} proposed an oscillatory exchange constant explicitly following the Ruderman-Kittel-Kasuya-Yosida (RKKY) formula. Assuming a high charge carrier concentration in the nanoclusters, this could lead to the competing interactions between the localized Mn spins inside the clusters required to form a spin glass state.

As mentioned above, the appearance of a concave shoulder in $M(T)$ curves indicates the presence of superparamagnetism in the samples. Li \textit{et al.} explained this behavior in the picture of BMPs.\cite{Li2005} These are formed around $T_C^*$ and grow in size as the temperature is lowered. In the limit of high Mn concentration, Kaminski \textit{et al.} predict that a BMP system undergoes a transition into a randomly ordered state in contrast to a ferromagnetic percolation transition.\cite{Kaminski2004} Indeed, glassy behavior has already been observed experimentally in Te compensated ${\rm Ga}_{0.915}{\rm Mn}_{0.085}{\rm As}$.\cite{Yuldashev2004} However, it is questionable, whether $\rm Ge_{0.96}\rm Mn_{0.04}$ can be described within the high Mn concentration limit of Ref.~\cite{Kaminski2004}.

The interpretation of the magnetization data by an onset of local ferromagnetism below a first transition temperature and the transition to a frozen, glassy state at a lower temperature due to cluster freezing is similar to the scenario reported for a so-called cluster glass material. A cluster glass consists of ferromagnetic clusters, which grow in size with decreasing temperature down to a temperature, at which they freeze due to intercluster frustration.\cite{Freitas2001} Like in the model of BMPs, at first local ferromagnetism occurs (the formation of ferromagnetic clusters in the cluster glass on the one hand and the formation of BMPs in the BMP model for DMSs on the other hand). These local ferromagnetic regions both grow in size, finally leading to a disordered glassy state at low temperatures.

In cluster glass materials, a two-peak structure in the susceptibility measurements was observed.\cite{Freitas2001} The peak in the susceptibility occurring at the higher temperature was assigned to be an indication for the formation of ferromagnetic clusters, whereas the low temperature peak was attributed to cluster freezing in the sample.\cite{Freitas2001} The temperature of the AC susceptibility peak of $\rm Ge_{0.8}\rm Mn_{0.2}$ ($\approx 115~\rm K$) is indeed close to the value of $T_C^*=128~\rm K$, determined from the temperature below which the formation of BMPs is supposed to set in following Li \textit{et al.} Therefore, in analogy to the cluster glass described above, the peak around $\approx115~\rm K$ in the AC susceptibility  might be connected to the onset of local ferromagnetism due to the formation of BMPs. However, the position of the high temperature peak observed by Freitas \textit{et al.},\cite{Freitas2001} which is thought to correspond to the local onset of ferromagnetism within the clusters, was found to be independent of frequency in contrast to the weak frequency dependence detected here. Therefore, the AC behavior exhibited by the Ge:Mn samples studied here is not completely identical to that reported for $\rm La_{0.7-x}\rm Y_x \rm Ca_{0.3}\rm MnO_3$ in Ref.~\cite{Freitas2001}.

\section{\label{sec:conclusion}CONCLUSION}

We have extensively studied the magnetic properties of \Ge-Mn with a focus on the low temperature state using three different methods spanning seven orders of magnitude in time scales. Instead of a ferromagnetic percolation transition,\cite{Li2005, Li2005a} we clearly find a glassy state below $T_f=12$~K {and $T_f=15$~K, for $x=0.04$ and $x=0.2$, respectively, only slightly depending on the Mn concentration}. For both samples, we observe a strong difference between the zero-field cooled and field cooled magnetization below $T_f$, as well as relaxation effects of the magnetization after switching the magnitude of the external magnetic field below $T_f$. In addition, AC susceptibility measurements on $\rm Ge_{0.96}\rm Mn_{0.04}$ show a peak around $T_f$, with the peak position $T'_f$ shifting as a function of the driving frequency $f$ by $\Delta T_f'/[T_f'\cdot \Delta \text{log} f]\approx 0.06$. {The more complicated behavior of AC susceptibility of sample $\rm Ge_{0.8}\rm Mn_{0.2}$ can be explained by a combination of ferromagnetic transition temperature of nanoclusters and their frequency dependent freezing.} These findings consistently show that Ge:Mn exhibits a frozen state at low temperatures, and that this dilute magnetic semiconductor can not be regarded as a conventional ferromagnet. {The spin glass-like magnetic behavior can be explained by a blocking transition of interacting superparamagnetic Ge:Mn nanoclusters at $T_f$.}

\begin{acknowledgments}
Our work was supported by Deutsche Forschungsgemeinschaft through SFB 631. Discussions with Stefan Ahlers, Dominique Bougeard, Martin Stutzmann{, and John A. Mydosh} are gratefully acknowledged.
\end{acknowledgments}

\end{document}